\documentclass[aps,superscriptaddress,twocolumn,]{revtex4}
\usepackage{bm}
\usepackage{epsfig}
\usepackage{bbm}
\usepackage{times}
\usepackage{amssymb}
\usepackage{amsmath}
\usepackage{natbib}
\usepackage{color}
\usepackage{soul}    
\usepackage[latin1]{inputenc}
\usepackage[loose,nice]{units}

\begin{document}
\title{Relativistic ionization dynamics for a hydrogen atom exposed to super-intense XUV laser pulses}

\author{Tor Kjellsson}
\affiliation{Department of Physics, Stockholm University, AlbaNova University Center, SE-106 91 Stockholm, Sweden}

\author{S{\o}lve Selst{\o}}
\affiliation{Faculty of Technology, Art and Design, Oslo and Akershus University College of Applied Sciences, NO-0130 Oslo, Norway}

\author{Eva Lindroth}
\affiliation{Department of Physics, Stockholm University, AlbaNova University Center, SE-106 91 Stockholm, Sweden}

\begin{abstract}
We present a theoretical study of the ionization dynamics of a hydrogen atom exposed to attosecond laser pulses in the extreme ultra violet region at very high intensities. The pulses are such that the electron is expected to reach relativistic velocities, thus necessitating a fully relativistic treatment. We solve the time dependent Dirac equation and compare its predictions with those of the corresponding non-relativistic Schr{\"o}dinger equation. We find that as the electron is expected to reach about 20 \% of the speed of light, relativistic corrections introduces a finite yet small decrease in the probability of ionizing the atom.
\end{abstract}

\maketitle

% pacs

% % % % % % % % % % % % % % % % % % % % %
% % % % % % Introduction  % % % % % % % %
% % % % % % % % % % % % % % % % % % % % % 
\section{Introduction}
\label{Introduction}
There are a number of infrastructure projects worldwide that strive for higher laser intensities, see e.g., the review in
Ref.~\cite{RevModPhysDiPiazza}. For lasers operating with long wavelengths, the intensities have already reached the regime
where magnetic interactions play a crucial role and ionized electrons move with relativistic velocities, see e.g., Refs.~\cite{Moore1999,DiChiara:08}.
With the latest generation of free electron lasers (XFEL, SACLA, LCLS), an unprecedented brilliance in the extreme ultraviolet (XUV) region and beyond is reached, and new techniques~\cite{Yumoto:13} to focus the beam, as well as preliminary results~\cite{Yoneda:14}, promise intensities also in this wavelength region exceeding $10^{20}$~W/cm$^2$. The treatment of light-matter interaction in a relativistic framework is thus a timely issue. To this end, some technical obstacles must be overcome -- obstacles specific to the relativistic time-dependent Dirac equation (TDDE). One issue is how to deal with the negative energy part of the spectrum of the Dirac Hamiltonian. Specifically, the stiffness induced by the huge energy difference between the positive and negative part of the spectrum may cause severe problems in resolving the dynamics. An even more challenging issue is, as it turns out, the consistent inclusion of higher order multipoles of the full electromagnetic field.

In Ref.~\cite{Selsto2009} the TDDE for hydrogen-like systems exposed to strong attosecond laser pulses was solved numerically. Unfortunately, computational constraints did not allow for calculations penetrating into the relativistic regime for hydrogen. It was found, however, that even below relativistic velocities, the inclusion of the negative energy part of the spectrum is crucial in order to account for dynamics beyond the dipole approximation. Moreover, in Ref.~\cite{Vanne2012} it was shown that even within the dipole approximation, negative energy states are crucial for ionization processes involving more than one photon.

More recently, it was demonstrated in Ref.~\cite{simonsen:16} that for a laser pulse with photon energy in the XUV-region and field strengths below the relativistic regime, higher order multipole effects are well accounted for by using the so called \textit{envelope approximation}, which does not contain the spatial dependence of the electromagnetic field in full. It was further demonstrated that within this approximation, the solutions of the TDDE and the non-relativistic time dependent Schr{\"o}dinger equation (TDSE) were in agreement.

In this work we go further and solve the equations of motion in the relativistic regime including multipole effects from the full field. To handle the computational load, which is quite heavy even for hydrogen, highly optimized parallel applications have been developed.

With these we investigate to what extent the ionization probabilities predicted by the TDDE differ from those of the TDSE, in a regime where relativistic corrections are to be expected.

Of course, relativistic effects arise when the electron is accelerated to velocities comparable to the speed of light. This may come about in two ways; for highly charged nuclei high velocities may be induced by the Coulomb potential alone.
Alternatively a strong external electromagnetic field can drive electrons towards relativistic speeds. In the former case, relativistic corrections to the energy structure do of course influence the ionization dynamics, e.g., by modifying the ionization potential \cite{Pindzola2010,Bauke2011,Vanne2012}. However, in the present work we will restrict ourselves to hydrogen and investigate cases in which the external field alone is strong enough to potentially induce relativistic dynamics.
As a ``measure of relativity'' we may take the maximum quiver velocity of a classical free electron exposed to a homogenous electric field of strength $E_0$ oscillating with frequency $\omega$:
\begin{equation}
\label{QuiverVel}
v_\mathrm{quiv} = \frac{e E_0}{m \omega} \quad .
\end{equation}
As $v_\mathrm{quiv}$ becomes comparable to the speed of light $c$, relativistic effects are expected.
In order to actually see such effects in the ionization probability, the laser specifications must, of course, be such that saturation is avoided even in this limit. Thus, the calculations will involve photon energies well into the XUV region. Various techniques have been applied in order to solve TDDE numerically, e.g., split operator methods -- combined with Fourier transforms \cite{Mocken2008,Bauke2011} or the method of characteristics \cite{Fillion-Gourdeau2012,Fillion-Gourdeau2014}, the close coupling method \cite{Pindzola2010,Pindzola2012} and Krylov methods \cite{Beerwerth2015}. In the present work TDDE is solved within a spectral basis as in Refs.~\cite{Selsto2009,Vanne2012}, i.e., the state is expanded in a set of eigenstates of the Hamiltonian without any external electromagnetic field present. The time propagation is performed using a low order Magnus expansion \cite{Blanes2009} while the actual matrix exponentiation is approximated by a Krylov subspace approach as in Ref.~\cite{Beerwerth2015}.

This paper is structured as follows: The next section outlines the theoretical framework, and details on the implementation are provided in Sec.~\ref{Implementation}. Our results and findings are presented and discussed in Sec.~\ref{ResAndDisc}, while our conclusions are drawn in Sec.~\ref{Conclusion}.
Atomic units are used throughout the text unless explicitly stated otherwise.

% % % % % % % % % % % % % % % % % % % % %
% % % % % %    Theory     % % % % % % % %
% % % % % % % % % % % % % % % % % % % % % 

\section{Theory}
\label{Theory}
Our starting point is the time-dependent Dirac equation (TDDE):
\begin{equation}
\label{TDDE}
i \hbar \frac{\mathrm{d}}{\mathrm{d} t} \Psi = H(t)\Psi,
\end{equation}
with the Hamiltonian
\begin{align}
\nonumber
H(t) & = c \boldsymbol{\alpha} \cdot \left[ {\bf p} +  e{\bf A}(\eta) \right] + V(r) \mathbbm{1}_4 + m c^2 \beta \\
\label{Hamiltonian}
&
= H_0 + e c\boldsymbol{\alpha} \cdot {\bf A} \quad .
\end{align}For the representation of $\boldsymbol{\alpha}$, the Pauli matrices are used,
\begin{equation}
\label{AlphaMat}
\boldsymbol{\alpha}
= \left( \begin{array}{cc} 0 &
\boldsymbol{\sigma}
\\
\boldsymbol{\sigma}
&  0 \end{array}\right) \quad ,
\end{equation}
and
\begin{equation}
\label{BetaMat}
\beta = \left( \begin{array}{cc} \mathbbm{1}_2 & 0 \\  0 &  -\mathbbm{1}_2 \end{array} \right) \quad .
\end{equation}
The four-component wave function can be written as
\begin{equation}
\label{EigenStatesOfH}
\Psi({\bf r},t)
=
\left( \begin{array}{c} \Psi_F({\bf r},t) \\ \Psi_G({\bf r},t)\end{array} \right) \quad  ,
\end{equation}
where $\Psi_F$ and $\Psi_G$ are two-component spinors.
The potential $V(r)$ is simply the Coulomb potential of a point nucleus, i.e., we neglect retardation effects in the electron-nucleus interaction and take the nuclear mass to be infinite, thus allowing for separation between the electronic and the nuclear degrees of freedom. The mass energy term, i.e., $m c^2 \beta$, introduces a $2m c^2$ gap in the spectrum, dividing it into the aforementioned negative and positive parts. Since changes in the population of negative energy states is interpreted as the appearance of positrons (through pair-creation), one might argue that the negative spectrum should be excluded in simulating the dynamics induced between an electron and an external field with strength far below the limit of pair production. This was disproved, however, in Refs.~\cite{Selsto2009,Vanne2012} and we will briefly return to this issue also in this work.

We choose to work in Coulomb gauge, $\nabla \cdot \bf{A} = 0$, with the external vector potential ${\bf A}$ linearly polarized along the $z$-axis and propagating along the $x$-axis;
\begin{equation}
\label{Adef}
{\bf A}(\eta) =  \frac{E_0}{\omega} f(\eta) \sin(\omega \eta + \varphi) \, \hat{\bf z}\quad ,
\end{equation}
where $\eta = t - x/c$. The envelope function is chosen to be sine-squared;
\begin{equation}
\label{EnvelopeDef}
f(\eta) = \left\{ \begin{array}{lc} \sin^2 \left( \frac{\pi \eta}{T} \right), & 0 < \eta < T  \\ 0, & \text{otherwise} \end{array} \right. \quad .
\end{equation}

With the pulse being linearly polarized in the $z$-direction,
the time-dependent part of the Hamiltonian becomes

\begin{equation} 
H_I (t) = c \alpha_z  A(x,t) \quad . 
\label{Hint}
\end{equation}

As this term depends on both time and space, $t$ and $x$, a direct calculation of the \textit{x}-dependent couplings induced by this interaction would have to be performed at each and every time step in order to represent $H(t)$ numerically. This cumbersome feature may be removed by writing the vector potential as a sum of terms with a purely time-dependent and a spatially dependent part,
\begin{equation}
\label{SeparateAinXT}
A(\eta) \approx \sum_{n=0}^{n_\mathrm{trunc}} c_n T_n(t) X_n(x) \quad .
\end{equation}
Such separations may be achieved by, e.g., a Fourier expansion in $\eta$ or a Taylor expansion around $\eta=t$. In Ref.~\cite{Selsto2009} both these approaches were followed. In the Fourier implementation, the number of terms was minimized in two ways: First, by taking $A$ to have the pulse length $T$ as period and, second, by neglecting the spatial dependence of the envelope $f(\eta)$, c.f., Eq.~(\ref{EnvelopeDef}). Both of these approaches have severe shortcomings. The former obviously introduces an erroneous periodicity, while for the latter it has been shown that in general it is the spatial dependence of the {\it envelope}, not the carrier, that provides the dominant correction to the dipole approximation~\cite{simonsen:16}.
In view of this we resort to a Taylor expansion in the present work:
\begin{equation}
\label{TaylorExpansion}
A(\eta) \approx \sum_{n=0}^{n_\mathrm{trunc}} \frac{1}{n!} A^{(n)}(t) \left( -\frac{x}{c} \right)^n
=
\sum_{n=0}^{n_\mathrm{trunc}} a_n(t) \, x^n
\end{equation}
without neglecting spatial dependence in neither the envelope nor the carrier.

The electric dipole approximation, which is generally not applicable for the cases of interest here, see, e.g., the discussion by Reiss~\cite{Reiss2000}, consists in substituting $\eta$ with the time $t$, i.e., neglecting the spatial dependence of the laser pulse completely. In going beyond the dipole approximation, i.e., assigning a value larger than zero to $n_\mathrm{trunc}$ in Eq.~(\ref{TaylorExpansion}), it is however important that the spatial dependence is introduced consistently in the equations of motion. This is not guaranteed by just choosing $n_\mathrm{trunc}=1$, which may be explained by first looking into the non-relativistic interaction.

% % % % % % % % % % % % % % % % % % % % % % % % % % % %
% % % % % %    Theory, subsection 1     % % % % % % % %
% % % % % % % % % % % % % % % % % % % % % % % % % % % %

\subsection{Non-relativistic interaction}
\label{NonRel_Interaction}
The time-dependent Schr{\"o}dinger equation is given by:
\begin{equation}
\label{TDSE}
i \hbar \frac{\mathrm{d}}{\mathrm{d} t} \Psi_\mathrm{NR} = H_\mathrm{NR}(t) \Psi_\mathrm{NR}
\end{equation}

with the Hamiltonian
\begin{equation}
\nonumber
H_\mathrm{NR}(t)  = \left[ \frac{p^2}{2m} + V(r) + \frac{e}{m}{\bf p} \cdot {\bf A} + \frac{e^2 A^2}{2m} \right].
\label{Hamiltonian_TDSE}
\end{equation}
In solving the TDSE, Eq.~(\ref{TDSE}), it has been found that the $A^2$-term, the so called {\it diamagnetic term}, provides practically all corrections to the dipole approximation \cite{Meharg2005,Forre2014}. Moreover, in Ref.~\cite{Forre2014} it was found sufficient to include only the first order correction in the diamagnetic term. Here it is worth emphasizing that in the case of the Schr\"odinger equation it is crucial that the {\it Hamiltonian}, thus {\it not} the vector potential itself, is expanded consistently in $x$.
Our resulting non-relativistic first order Hamiltonian is then;
\begin{equation}
\label{H_NR_BYD1} H_\mathrm{NR} \approx \frac{p^2}{2m}+ V(r) +
\frac{e}{m} p_z A(t) -  \frac{e^2}{m} \frac{x}{c} A(t) A^{(1)}(t) \quad ,
\end{equation}
where the purely time-dependent $A(t)^2$-term has been removed by the trivial gauge transformation:
\begin{equation}  \tilde{\Psi}(\mathbf{r},t) = e^{\frac{i}{\hbar} \int_{0}^{t} \frac{e^2}{2m} A(\omega t')^2 dt' }\Psi(\mathbf{r},t). \end{equation}

In the non-relativistic regime the TDSE and the TDDE should, of course, agree. This condition will now prove useful in understanding the consistent incorporation of effects beyond the dipole approximation in the Dirac Hamiltonian by studying its non-relativistic limit.

% % % % % % % % % % % % % % % % % % % % % % % % % % % %
% % % % % %    Theory, subsection 2     % % % % % % % %
% % % % % % % % % % % % % % % % % % % % % % % % % % % %

\subsection{The non-relativitic limit of the light-matter interaction }
\label{nonrellimit}

While the time dependent Schr{\"o}dinger equation, Eq.~(\ref{TDSE}), has both a linear and a quadratic term in ${\bf A}$, the time dependent Dirac equation, Eqs.~(\ref{TDDE},\ref{Hamiltonian}), is only linear in the vector potential. We can study the non-relativistic limit of the TDDE by using
the form of the wave function given in Eq.~(\ref{EigenStatesOfH}) and rewrite Eq.~(\ref{TDDE}):
\begin{align}
& V \Psi_F +c \boldsymbol{\sigma} \cdot \left( e \mathbf{A} + \mathbf{p} \right) \, \Psi_G = i\hbar \frac{\mathrm{d} \Psi_F}{\mathrm{d} t}
\label{twokomp1}
%\nonumber
\\
& c \boldsymbol{\sigma} \cdot \left(  e \mathbf{A} + \mathbf{p} \right)   \Psi_F
+ \left( V - 2mc^2 \right) \, \Psi_G = i\hbar
\frac{\mathrm{d} \Psi_G}{\mathrm{d} t}
 \quad ,
\label{twokomp}
\end{align}
with $\Psi_F$ and $\Psi_G$ being the upper and lower component, respectively, c.f. Eq.~(\ref{EigenStatesOfH}). For positive energy states,
 \begin{equation*} |\Psi_F| \sim c \, |\Psi_G| \quad ,\end{equation*}
so a comparison with $\Psi_\mathrm{NR}$ should be dictated by $\Psi_F$.

By assuming that the Coulomb potential is negligible in comparison with the mass energy term, $V \ll 2mc^2$, and that the time variation of the small component $\Psi_G$ is modest, Eq.~(\ref{twokomp}) yields
\begin{align}
\Psi_G \approx  \frac{1}{2mc} \boldsymbol{\sigma} \cdot \left(  e \mathbf{A} + \mathbf{p} \right)   \Psi_F \quad ,
\end{align}
which inserted into Eq.~(\ref{twokomp1}) provides
\begin{equation}
\label{nonrellimit_eq}
\left[ \frac{p^2}{2m} + V +\frac{e}{m} \mathbf{p} \cdot \mathbf{A} + \frac{e^2 A^2}{2m} \,
+ \frac{e \hbar}{2 m}  \boldsymbol{\sigma}
\cdot \mathbf{B} \right] \Psi_F
= i\hbar \frac{\mathrm{d} \Psi_F}{\mathrm{d} t}.
\end{equation}
That is, the TDSE, Eq.~(\ref{TDSE}), is reproduced -- with an additional term corresponding to the interaction between the spin and the magnetic field.
The same can be achieved via a Foldy-Wouthuysen type of transformation on the wave function~\cite{foldy:50}.

Specifically, the $A^2$-term seen in the non-relativistic Hamiltonian of Eq.~(\ref{TDSE}) reappears.
Thus, we see that the $A^2$-term is implicitly present in the Dirac equation and enters the equation for the large component of the wave function, $\Psi_F$, through the small component, $\Psi_G$. Whenever we try to formulate the interaction with the electromagnetic field through operators that are anti-diagonal with respect to the small and large component, it reappears. This implicit occurence complicates a consistent inclusion of spatial effects in the relativistic interaction, as will be demonstrated in Sec.~\ref{ResAndDisc}.

% % % % % % % % % % % % % % % % % % % % % % % % % % % %
% % % % % %    Theory, subsection 3     % % % % % % % %
% % % % % % % % % % % % % % % % % % % % % % % % % % % %

\subsection{Propagation}
\label{PropAndProj}
Both in the relativistic and the non-relativistic case, the state vector $\Psi(t)$ is propagated by means of a second order Magnus propagator,
\begin{equation}
\label{Propagator}
\Psi(t+\tau) = \exp[-i \tau H(t+\tau/2)] \Psi(t) + \mathcal{O}(\tau^3)\quad .
\end{equation}
One of the major advantages of a Magnus-type propagator for the Schr{\"o}dinger equation is stated clearly in Ref.~\cite{Hochbruck2003}:
``{\it In
contrast to standard integrators, the error does not depend on higher time derivatives of the solution,
which is in general highly oscillatory}''.
Due to the stiffness inherent in the mass energy term, i.e., the $mc^2 \beta$-term of the Hamiltonian in Eq.~(\ref{Hamiltonian}),
this becomes even more advantageous for the Dirac equation than for the Schr{\"o}dinger equation. In fact, the accuracy of many time-propagation schemes, such as Crank-Nicolson and Runge-Kutta, suffer greatly from the $2 m c^2$ energy splitting. In Ref.~\cite{Salamin2006}, e.g., it is stated that ``{\it
The major drawback of the Dirac treatment is the temporal step size $\Delta t \lesssim \hbar/E$ required, which has to be significantly
smaller than for Schr{\"o}dinger treatments, because of the large rest mass energy $mc^2$ that is contained in the particle's
total energy E.}''
Indeed, several works dealing numerically with the TDDE apply time-steps of the order $10^{-5}$~a.u. or smaller, see, e.g., \cite{Mocken2003,Pindzola2010,Bauke2011,Fillion-Gourdeau2014}. Obviously, such a restriction renders the description of a laser pulse of a realistic duration rather infeasible.
This problem is, however, circumvented by the application of Magnus propagators~\cite{Hochbruck2003}.
It should be noted that extremely small time steps are, of course, required when resolving phenomena which really do take place at such short time scales, such as Zitterbewegung~\cite{Beerwerth2015}.

We emphasize that in a time-dependent context it is crucial to keep in mind that a time dependent Hamiltonian leads to a time-dependent distinction between positive and negative energy states; the Dirac sea is not calm, so to speak.
As was discussed in Ref.~\cite{Selsto2009}, negative energy solutions, as defined by the time-independent Hamiltonian, $H_0$, are essential to calculate effects beyond the dipole approximations, while the dynamic negative energy states, as defined by $H(t)$, should only come into play when pair-production starts to play a role. For fields far away from that limit
the propagator of Eq.~(\ref{Propagator}) may in principle be modified to
\begin{equation}
\label{PropagatorProj}
\Psi(t+\tau) = \mathcal{P}(t+\tau/2) \exp[-i \tau H(t+\tau/2)]  \Psi(t) + \mathcal{O}(t^3)\quad ,
\end{equation}
where $\mathcal{P}(t)$ projects the state onto the time-dependent subspace spanned by the positive spectrum of the Hamiltonian $H(t)$, i.e., the negative energy states are blocked. For the fields used here the population of the  negative energy states of the Hamiltonian $H(t)$ is so tiny that the projected and the unprojected propagator gives the same results, as 
explained in Sec.~\ref{ResAndDisc}.

When, ultimately, the fields are increased even further and pair-production do start to play a role, this has, of course, to be handled through field theory, where the distinction between positive and negative energy states is inherent. Then transitions into negative energy states (annihilation) are coupled to  excitations out of them (pair-production) and the number of particles is no longer conserved.

% % % % % % % % % % % % % % % % % % % % % % % % %
% % % % % %    Implementation     % % % % % % % %
% % % % % % % % % % % % % % % % % % % % % % % % % 

\section{Implementation}
\label{Implementation}
We expand our wave function in eigenstates of the unperturbed Hamiltonian $H_0$,
\begin{equation}
\label{Expansion}
\Psi(t) = \sum_{n,j,m,\kappa} c_{n,j,m,\kappa}\left( t \right) \psi_{n,j,m,\kappa}({\bf r})  \quad ,
\end{equation}
with
\begin{equation}
\label{EigenStatesOfH0}
\psi_{n,j,m,\kappa}({\bf r})
=
\left( \begin{array}{c} F_{n,j,m,\kappa}({\bf r}) \\ G_{n,j,m,\kappa}({\bf r})\end{array} \right) \quad  ,
\end{equation}
where
\begin{equation}
\label{EigenStatesOfH0_radial}
\left( \begin{array}{c} F_{n,j,m,\kappa}({\bf r}) \\ G_{n,j,m,\kappa}({\bf r})\end{array} \right)
 = \frac{1}{r} \left( \begin{array}{c} P_{n,\kappa}(r) X_{\kappa,j,m}(\Omega) \\ i Q_{n,\kappa}(r) X_{\textrm{-}\kappa,j,m}(\Omega) \end{array} \right) \quad .
\end{equation}
Here $\kappa=l$ for $j=l-1/2$ and $\kappa=-(l+1)$ for $j=l+1/2$, and $X_{\kappa,j,m}$ represents the spin-angular part which has the analytical form
\begin{equation} \mathit{X}_{\kappa, j, m} = \sum_{m_s,m_l} \langle l_{\kappa},m_l ;s,m_s | j,m \rangle {Y}^{l_{\kappa}}_{m_l} (\theta, \phi) \chi_{m_s} \quad ,
\end{equation}

where ${Y}^{l_{\kappa}}_{m_l} (\theta, \phi)$ is a spherical harmonic and $\chi_{m_s}$ is an eigenspinor. The radial components $P_{n,\kappa}(r)$ and $Q_{n,\kappa}(r)$ are expanded in B-splines~\cite{Boor1978};
\begin{equation} \label{P_Q_Bspl} P_{n,\kappa}(r) = \sum_{i} a_i B_i^{k_1}(r) \quad, \quad
Q_{n,\kappa}(r) = \sum_{j} b_j B_j^{k_2}(r) \quad .
\end{equation}

In Ref.~\cite{Charlotte2009} it is demonstrated that specific
choices of B-spline orders $k_1$ and $k_2$ control the occurrence of the so called \textit{spurious states}, which are known to appear in the
numerical spectrum after discretization of the Dirac Hamiltonian.
While the choice $k_1=k_2$ contaminates the spectrum with such incorrect states, the choices $k_1=k_2 \pm 1$ are reported to be stable combinations that do not produce them. 

In this work we use $k_1=7, k_2 = 8$. Converged results were obtained using a linear knot sequence with 500 B-splines for the large component and 501 for the small one up to $R_\mathrm{max} = 150$~a.u.. This gives a total of 1001 bound and pseudo continuum (both positive and negative) states that the energy index $n$ can attain per spin-angular symmetry. The boundary conditions applied on the components are:
\begin{eqnarray}  P_{n,\kappa}(0) & = & P_{n,\kappa}(R_\mathrm{max}) = 0  \\
 Q_{n,\kappa}(0) & = & Q_{n,\kappa}(R_\mathrm{max}) = 0 \quad . \end{eqnarray}

We include all spin-orbitals with angular momentum up to $l_\mathrm{max} = 30$ (as defined for the large component) and keep all the associated magnetic quantum numbers $m_j$. To speed up the propagation without compromising the results, high energy components have been filtered out leaving, for this typical choice of parameters a final number of $1\,902\,594$ states in our basis. Similarly, the non-relativistic spectral basis has eigenfunctions of the form
\begin{equation} \Phi({\bf r}) =  \frac{P_{nl}({r})}{r} Y_m^l(\theta,\phi) \quad ,
\end{equation}
where, as in the relativistic case, the radial component is expanded in B-splines while the angular part is analytically known. We use the same linear knot sequence as in the relativistic case with 500 B-splines, $k=7$ and $R_\mathrm{max}=150$~a.u. and the boundary conditions,
\begin{equation}
P_{n,l}(0) = P_{n,l}(R_\mathrm{max}) = 0 \quad .
\end{equation}
For convergence we needed to keep all orbital angular momenta up to $l_\mathrm{max} = 40$, with all associated $m_l$-values. Just as in the relativistic case we filtered out high energy components that do not affect the dynamics, leaving a total of $827\,351$ states in the non-relativistic basis.

% % % % % % % % % % % % % % % % % % % % % % % % % % % % % % %
% % % % % %    Implementation - subsection 1  % % % % % % % %
% % % % % % % % % % % % % % % % % % % % % % % % % % % % % % % 
\subsection{Computing the matrix elements}
\label{Interaction_elements}

With the vector field given by Eq.~(\ref{TaylorExpansion}), the light-matter interaction, Eq.~(\ref{Hint}), gives rise to matrix elements between eigenstates to the time-independent Hamiltonian. 
Labelling two such states $ | k \rangle = | n \kappa j m \rangle $  and $ | \tilde{k} \rangle = | \tilde{n} \tilde{\kappa} \tilde{j} \tilde{m}\rangle$, the matrix element connecting them is given by a sum of terms including higher and higher powers of the spatial coordinate:
\begin{equation}
\label{H_BB} H_{k\tilde{k}}(t) = \sum_{\gamma=0}^{n_{\rm trunc}} a_i(t) \langle n \kappa j m  | \alpha_z x^{\gamma}  | \tilde{n} \tilde{\kappa} \tilde{j} \tilde{m}\rangle. \end{equation}
Since we are working in the eigenbasis of $H_0$, we may compute these couplings using angular momentum theory by first expressing the operators in terms of spherical tensor operators~\cite{mbpt}. For the powers of $x$, these are spherical harmonics $Y^{\lambda}_{\mu}$;
\begin{equation}
x^{\gamma} = \sum_{\lambda=0}^{\gamma} \sum_{\mu=-\lambda}^{\lambda}
c_{\gamma}^{\lambda \mu} r^{\gamma} Y^{\lambda}_{\mu} \quad ,
\end{equation}
while $\sigma_z = \sigma^1_0$ is a component of a rank-1 spherical tensor operator. 
The couplings in Eq.~(\ref{H_BB}) are now given by summing up terms factored in a radial and spin-angular part:
\begin{align}
\label{mat_el_radial}
& \langle n \kappa j m  | \alpha_z r^\gamma Y^{\lambda}_{\mu}  | \tilde{n} \tilde{\kappa} \tilde{j} \tilde{m}\rangle =
\nonumber \\
&
i \int_{r} \Big[ P^*_{n \kappa}(r) r^{\gamma} Q_{\tilde{n} \tilde{\kappa}}(r) \langle \kappa j m | \sigma^1_0 Y^{\lambda}_{\mu} |\textrm{-} \tilde{\kappa} \tilde{j} \tilde{m} \rangle -
\nonumber \\
& Q^*_{n \kappa}(r) r^{\gamma} P_{ \tilde{n} \tilde{\kappa} }(r) \langle \textrm{-} \kappa j m | \sigma^1_0 Y^{\lambda}_{\mu}  | \tilde{\kappa} \tilde{j} \tilde{m} \rangle \Big] \, dr
\end{align}
where $\textrm{-} \kappa$ implies the spin-angular part of the small component. 
With the radial components expressed in B-splines, the integrals are computed to machine accuracy using Gauss-Legendre quadrature. To obtain the spin-angular part, the operator product can be expressed in a coupled tensor operator basis:
\begin{equation*} \sigma^1_0 Y^{\lambda}_{\mu} = \sum_{K=|\lambda -1|}^{\lambda+1} \langle 1 0 ; \lambda \mu | K Q \rangle \left\lbrace \pmb{\sigma}^1 \mathbf{Y}^{\lambda}\right\rbrace^K_Q \; ,
\quad Q = \mu \quad .
\end{equation*}

\noindent With this choice the spin-angular part may be computed as:
\begin{align}
 \langle  \kappa j m | \sigma^1_0 Y^{\lambda}_{\mu} |\tilde{\kappa} \tilde{j} \tilde{m} \rangle  =  \sum_{K=|\lambda -1|}^{\lambda+1} \left(
-1 \right)^{\lambda-1-Q} \sqrt{2K+1} \nonumber \\
 \times \begin{pmatrix}
1 & \lambda & K \\
0 & \mu & -\mu
\end{pmatrix}    \langle  \kappa j m | \left\lbrace\pmb{\sigma}^1 \mathbf{Y}^{\lambda}\right\rbrace^K_Q |\tilde{\kappa} \tilde{j} \tilde{m} \rangle \quad .
\end{align}
\noindent The Wigner-Eckart theorem can be applied to the matrix element of the combined operator $\left\lbrace\pmb{\sigma}^1 \mathbf{Y}^{\lambda}\right\rbrace^K_Q $:

\begin{align}
\label{Sigma_Sph_Harm_cpls}
&\langle  \kappa j m | \left\lbrace \pmb{\sigma}^1 \mathbf{Y}^{\lambda}\right\rbrace^K_Q |\tilde{\kappa} \tilde{j} \tilde{m} \rangle =
\nonumber \\
&\left(\textrm{-}1 \right)^{j-m}  \begin{pmatrix}
j & K & \tilde{j} \\
-m & Q & \tilde{m} \end{pmatrix} \times
\langle j || \left\lbrace\pmb{\sigma}^1 \mathbf{Y}^{\lambda}\right\rbrace^K  || \tilde{j} \rangle
\end{align}
\noindent with the reduced matrix element given by:
\begin{align} \label{combined_redmat} \langle j || \left\lbrace\pmb{\sigma}^1 \mathbf{Y}^{\lambda}\right\rbrace^K  || \tilde{j} \rangle  = \sqrt{(2j+1)(2K+1)(2\tilde{j}+1)}
\nonumber \\
\times
\langle s || \pmb{\sigma}^1 || \tilde{s} \rangle  \langle l || \mathbf{Y}^{\lambda} || \tilde{l} \rangle \begin{Bmatrix} l & \tilde{l} &  1 \\ s &    \tilde{s} &  \lambda \\ j & \tilde{J} & K \end{Bmatrix} \quad .
\end{align}

\noindent With this scheme the couplings induced by $\alpha_z x^{\gamma}$, for \ $\gamma=0,1, \ldots, n_\mathrm{trunc}$ in Eq.~(\ref{H_BB}) are readily computed to represent the light-matter interaction term, cf.~Eq.~(\ref{Hint}), in the  Dirac Hamiltonian, Eq.~(\ref{Hamiltonian}). For the Schr{\"o}dinger Hamiltonian in Eq.~(\ref{H_NR_BYD1}) the needed couplings are $p_z$-couplings;

\begin{align}
& \langle n_a l_a m_a | p_z |n_b l_b m_b \rangle =
\nonumber \\
&
 \left(-1 \right)^{l_a-m_a}
	\begin{pmatrix} 	 l_a  &  1  &  l_b \\
		 				-m_a  &  0  &  m_b  	
	\end{pmatrix}
\times	 \langle n_a l_a || \mathbf{p}^1 || n_b l_b\rangle
\quad ,
\end{align}
and $x$-couplings;
\begin{align}
&
\langle n_a l_a m_a | x | n_b l_b m_b \rangle =
\nonumber \\
&
 \label{x_mat_el} \sqrt{\frac{2 \pi}{3}} \cdot \left(-1 \right)^{l_a-m_a} 
 \left(	 \begin{pmatrix}  	 l_a  &  1  &  l_b \\
			 			   	-m_a  &  -1  &  m_b
		 \end{pmatrix}
			 			   	- 	 				
	     \begin{pmatrix}	 l_a  &  1  &  l_b \\
			 				-m_a  &  1  &  m_b
	     \end{pmatrix}	\right) \nonumber \\
&
 \times	\int P_{n_a l_a}^{*}(r) \  r \ P_{n_b l_b}(r) dr \ \langle n_a l_a || \mathbf{Y}^1 || n_b l_b\rangle \quad ,
\end{align}
where, just as in Eq.~(\ref{combined_redmat}), the Wigner-Eckart theorem has been applied. Note that this enables an efficient memory storage since all dependence on the projection numbers may be accounted for by multiplication of single scalars. This fact has also been used for cache-efficient memory usage in the propagation method used.

% % % % % % % % % % % % % % % % % % % % % % % % % % % % % % %
% % % % % %    Implementation - subsection 2  % % % % % % % %
% % % % % % % % % % % % % % % % % % % % % % % % % % % % % % % 
\subsection{Propagation}
As the use of a Magnus propagator, Eq.~(\ref{Propagator}), involves exponentiating a time-dependent Hamiltonian matrix, full diagonalization at each time step is called for. This would also be necessary in order to make the distinction between (dynamical) positive and negative energy states needed for the projection in Eq.~(\ref{PropagatorProj}). Incidentally, in the work of Ref.~\cite{Selsto2009}, imposing this projection was actually necessary due to the use of the complex scaling method, which was then a key to minimize the basis size. Since negative energy states attain a positive imaginary energy component under complex scaling, they cannot be propagated forward in time and has to be separated from the positive energy states.

Needless to say, repeated diagonalization of a double precision matrix of size \ $(10^6,10^6)$ \ is extremely expensive and completely out of the question. Any efficient way of doing the exponentiation approximatively is thus welcome.
To this end, Krylov subspace methods turn out to be quite useful \cite{Hochbruck1997, Beerwerth2015}. Such methods provide accurate approximations to the action of the exponential of an operator on a specific vector. Moreover, their numerical implementation can be made
very efficient. At each time step the Krylov subspace of dimension $m$, $K_m(t+\tau/2)$, which is spanned by
the set of states obtained by iteratively multiplying the state with the Hamiltonian,
\begin{equation}
\label{KrylovSpace}
\left[H(t+\tau/2)\right]^k \Psi(t), \quad k=0,...,(m-1) \quad ,
\end{equation}
is constructed using the Arnoldi algorithm. The exponential is now projected onto $K_m(t+\tau/2)$ and exponentiated within this subspace.
Typically $m \approx 50$ when using a time step of $\tau \approx 10^{-3}$~a.u. in our calculations. Once convergence is achieved, a back transformation to the original Hilbert space is performed to give $\Psi(t+\tau)$. The computationally heavy part in this approach is the repeated matrix-vector products in Eq.~(\ref{KrylovSpace}). Despite $H(t+\tau/2)$ being quite sparse, the number of non-zero elements in the relativistic case is approximately $3.9 \cdot 10^{11}$, which is quite a challenge to handle. This is done by storing the projection factors separate from the rest of the couplings, as discussed in Sec.~\ref{Interaction_elements}, such that the memory requirement for our parameters is reduced by roughly a factor $\sim 100$ while high computational throughput is achieved by performing all multiplications corresponding to transitions between states sharing all other quantum numbers simultaneously.

In order to prevent reflection at the computational box boundary, a complex absorbing potential (CAP) is added to $H(t)$ during the time-propagation -- for both the Schr\"odinger and the Dirac equation,
\begin{equation}
V_\mathrm{CAP} = \left\{\begin{array}{cc}
- \eta \left(r -r_0 \right)^2 , & r >r_0 \\
0, & r\leq r_0
\end{array}
\right. \quad .
\end{equation}
Typical CAP values are $\eta =0.05i$ and $r_0=110.0$~a.u., the latter being large enough to ensure that any flux reaching this distance will really represent the ionization current. In that case the absorption of the flux beyond $r_0$ does not affect the ionization dynamics -- provided that reflection is negligible. For the TDDE the complex absorbing potential is found to have the same effect on all the components of the wave function.

% % % % % % % % % % % % % % % % % % % % % % % % % % % % %
% % % % % %    Results and discussion     % % % % % % % %
% % % % % % % % % % % % % % % % % % % % % % % % % % % % % 

\section{Results and discussion}
\label{ResAndDisc}
The probatility of ionizing a hydrogen atom from the ground state, $P_\mathrm{ion}$, by exposing it to laser pulses of various intensities has been studied. The pulse is characterized by the following parameters, cf. Eqs.~(\ref{Adef},\ref{EnvelopeDef}):
\begin{equation}
\omega = 3.5~\mbox{a.u.},  \ \ \phi = 0, \ \ T = 2 \pi \tfrac{N_c}{\omega}~ \mbox{a.u.} \ \ \text{and} \ \ N_c = 15.
\end{equation}
$P_\mathrm{ion}$ has been calculated with peak electric field strengths,  $E_0$, so high that the electron quiver velocity, $v_\mathrm{quiv}$, corresponds to  almost 20 \% of $c$, cf. Eq.~(\ref{QuiverVel}). But before we discuss such extreme conditions we will consider the ionization probability in the non-relativistic regime.

\begin{figure}[h]
 \includegraphics[width=0.50\textwidth]{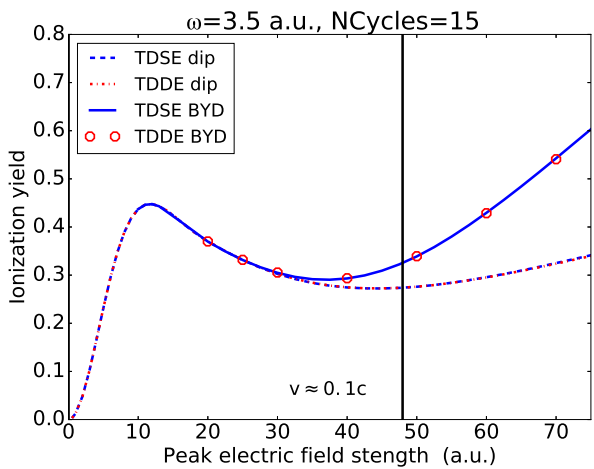}
\caption{(color online). Comparison of TDSE and TDDE calculations within the dipole approximation and beyond (BYD). The vertical line indicates a maximum electric field strength $E_0$ corresponding to a maximum quiver velocity $v_\mathrm{quiv} = 0.1c$.}
\label{Pion_low}
\end{figure}

Figure~\ref{Pion_low} shows the converged $P_\mathrm{ion}$ calculated both within the dipole approximation and beyond (BYD) for $E_0 \leq 70$~a.u. (corresponding to $I \sim 1.7 \cdot 10^{20} $ $\mbox{W/cm}^2$). For comparatively low field strengths the ionization probability is increasing
monotonously with increasing field strength, but around $E_0 \approx 10$~a.u. the so-called {\it stabilization} sets in~\cite{gavrila2002}. For even higher field strengths the ionization starts to increase again. At $E_0 \approx 30$~a.u. the dipole approximation breaks down, a behaviour that has been discussed, e.g., in Ref.~\cite{Simonsen-Forre::15}.

The vertical line in Fig.~\ref{Pion_low} marks the electric field strength $E_0$ that corresponds to a maximum electron quiver velocity $v_\mathrm{quiv} = 0.1c$. As expected, the predictions by the TDSE and the TDDE do indeed agree below and around this regime. However, the convergence patterns with respect to the spatial dependence in Eq.~(\ref{TaylorExpansion}) of the calculations are actually very different. We will now study this convergence behavior in some detail.

% % % % % % % % % % % % % % % % % % % % % % % % % % % % % % % % % % % %
% % % % % %    Results and discussion - subsection 1    % % % % % % % %
% % % % % % % % % % % % % % % % % % % % % % % % % % % % % % % % % % % %

\subsection{The representation of the vector field beyond the dipole approximation }

\begin{figure}[!h]
\includegraphics[width=0.50\textwidth]{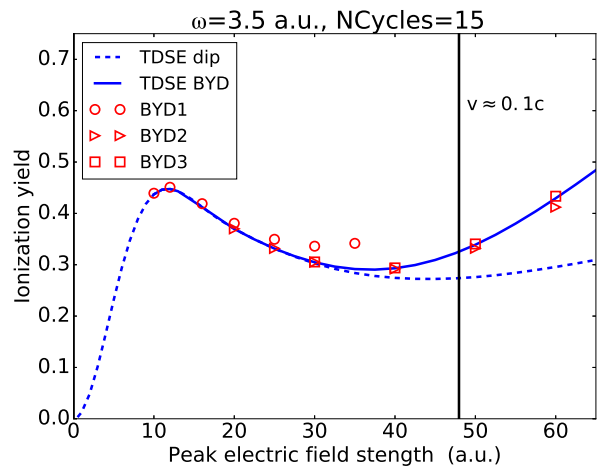}
\caption{(color online). Comparison between TDDE results with increasing order of $x$ included in the expansion of the vector potential. The numbers trailing the acronym ``BYD'' is the numerical value of $n_\mathrm{trunc}$ in Eq.~(\ref{TaylorExpansion}). BYD1 starts deviating from TDSE already at $E_0 \approx 12$~a.u. and from there on consistently overestimates $P_\mathrm{ion}$. BYD2, on the other hand, agrees with TDSE up to $E_0 = 40$~a.u. where it starts to give a \textit{lower} value for $P_\mathrm{ion}$. BYD3 then pushes $P_\mathrm{ion}$ up and agrees with the TDSE results up to $E_0 \approx 60$~a.u..}
\label{BYD1-3}
\end{figure}

As previously mentioned, Ref.~\cite{Forre2014} provides strong support for the claim that below the relativistic region, the first order term in Eq.~(\ref{TaylorExpansion}) alone provides practically all corrections to the dipole prediction. At first glance there does not seem to be any a priori reason why this conclusion should not apply to the relativistic treatment as well. Yet, as shown in Fig.~\ref{BYD1-3}, we typically have to resort to a third order expansion to reproduce the non-relativistic results when solving the Dirac equation.

The underlying problem is the implicit inclusion of the $A^2$-term in the Dirac-equation, discussed in Sec.\ref{nonrellimit}. When only {\em first} order corrections to $\mathbf{A}$ are included, some of the {\em second} order corrections in the {\em implicit} $A^2$-term will still be included, namely the $\sim (A^{(1)})^2 x^2$ contribution, while the other contribution, namely the $\sim A^{(2)}A^{(0)} x^2$-term, require the inclusion also of second order corrections to $\mathbf{A}$. F{\o}rre and Simonsen~\cite{Forre2014}
found that in the non-relativistic limit there are important cancellations between these $x^2$-terms; its net effect is vanishingly small  while separate contributions would shift results dramatically -- and artificially.

It is thus natural to assume that these cancellations are a key-issue and that higher-order correction terms to $\mathbf{A}$ are needed to achieve them when 
implementing a solution of the Dirac equation \cite{PrivateForre15}.
Of course, inclusion of second order correction terms in $A$, in turn, introduces effective third and fourth order terms in an equally inconsistent manner and so forth. However, since the magnitude of these corrections decrease with $n_\mathrm{trunc}$, cf. Eq.~(\ref{TaylorExpansion}), the problem should
for a given field strength diminish with increasing orders. This  convergence behavior is observable in Fig.~(\ref{BYD1-3}); TDDE BYD1, i.e., $n_\mathrm{trunc} = 1$, severely overestimates the ionization yield even before the breakdown of the dipole approximation, while TDDE BYD2, i.e., $n_\mathrm{trunc} = 2$, gives initial agreement up to $E_0 = 40$~a.u. but is then seen to \textit{underestimate} $P_\mathrm{ion}$ (as predicted by the TDSE). Inclusion of an additional third order term in Eq.~(\ref{TaylorExpansion}) increases $P_\mathrm{ion}$ as compared to TDDE BYD2, giving close agreement with the TDSE up to $E_0=60$~a.u.. 

Before we go on, a comment is necessary. The conclusion concerning the inconsistent representation of the implicit $A^2$-term may seem to contradict the results presented in Ref.~\cite{simonsen:16}, where the agreement was found between TDSE and TDDE using a \textit{first} order expansion in $x$ of ${\bf A}$ utilizing the so called envelope approximation. With this approximation the spatial dependence in the \textit{carrier}, i.e., $\sin(\omega \eta + \varphi)$ in Eq.~(\ref{Adef}), is disgarded, which for the TDDE in fact removes the inconsistency problem to a large extent: The remaining term in the derivative of the vector potential, $A^{(1)}(t)$, now only consists of a small contribution from the envelope, cf. Eq.~(\ref{EnvelopeDef}), and thus the problematic $\sim (A^{(1)})^2 x^2$ contribution from the implicit $A^2$ term in Eq.~(\ref{nonrellimit_eq}) is more or less insignificant -- at least in the non-relativistic regime. Within the envelope approximation it is thus sufficient to use only the first order term in the Taylor-expansion  of ${\bf A}$, but it is not when also its full spatial dependence is considered.

In order to find relativistic effects we now consider higher values for $E_0$.

% % % % % % % % % % % % % % % % % % % % % % % % % % % % % % % % % % % %
% % % % % %    Results and discussion - subsection 2    % % % % % % % %
% % % % % % % % % % % % % % % % % % % % % % % % % % % % % % % % % % % %

\subsection{Relativistic effects}

\begin{figure}[!h]
\includegraphics[width=0.50\textwidth]{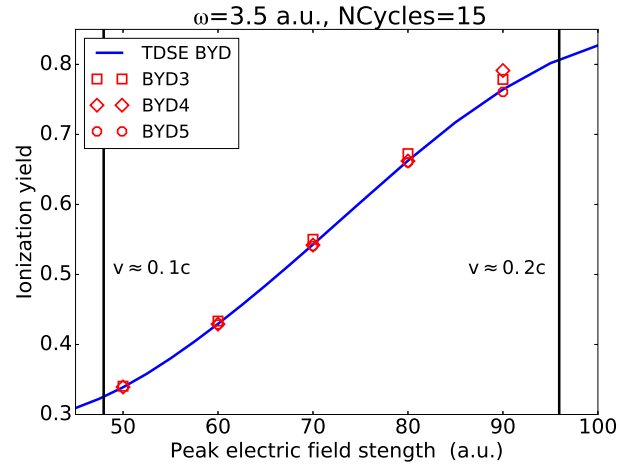}
\caption{(color online). Continuation of Fig.~\ref{BYD1-3} with $n_\mathrm{trunc}=3, 4$ and $5$ for the TDDE. At the highest field strengths 
the relativistic $P_\mathrm{ion}$ starts to show a decreasing value compared to the non-relativistic prediction. }
\label{BYD3-5}
\end{figure}

\begin{figure}[!h]
\includegraphics[width=0.50\textwidth]{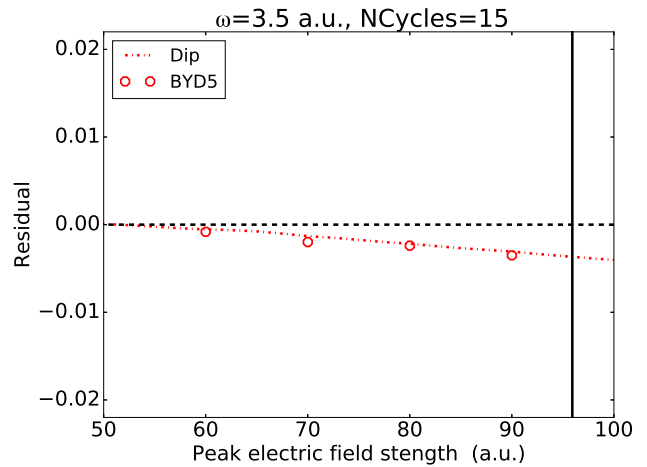}
\caption{(color online). Residual plot for the final converged TDDE calculations with respect to TDSE. Both within and beyond the dipole approximation the relativistic corrections show a decrease of $P_\mathrm{ion}$ as the quiver velocity $v_\mathrm{quiv}$ approaches $0.2c$.}
\label{Residual_plot}
\end{figure}

Figure~\ref{BYD3-5} shows $P_\mathrm{ion}$ for electric field strengths up to $E_0=90$ a.u. (corresponding to $I \sim 2.8 \cdot 10^{20} $ $\mbox{W/cm}^2$). 
For higher field strengths than those seen in Fig.~\ref{Pion_low}, higher values of $n_\mathrm{trunc}$ are necessary, as shown in Fig.~\ref{BYD3-5}.
For the highest field strengths considered here, relativistic effects start to surface. In order to highlight these we present the difference between the relativistic and the non-relativistic predictions in Fig.~(\ref{Residual_plot}). Both within and beyond the dipole approximation the relativistic $P_\mathrm{ion}$ seems to decrease steadily compared to the corresponding non-relativistic value. Albeit small, the discrepancy is well within the accuracy of our calculations.

It is interesting to note from Fig.~\ref{Residual_plot} that the difference between the relativistic and the non-relativistic treatment is more or less the same within the dipole approximation as beyond it. This indicates that, although the velocity of the electron in the polarization direction, $v_{\rm quiv}$,  cf.~Eq.~(\ref{QuiverVel}), induced by the electric field, is subject to relativistic corrections, the magnetic interaction is essentially unaffected. This is not surprising, however, considering the magnitude of the relativistic correction here and the fact that the  velocity in the direction perpendicular to the direction of polarization, induced by the magnetic field, is smaller by a factor $v_{\rm quiv}/c$, i.e., still far from relativistic.

\begin{figure}[!h]
\includegraphics[width=0.50\textwidth]{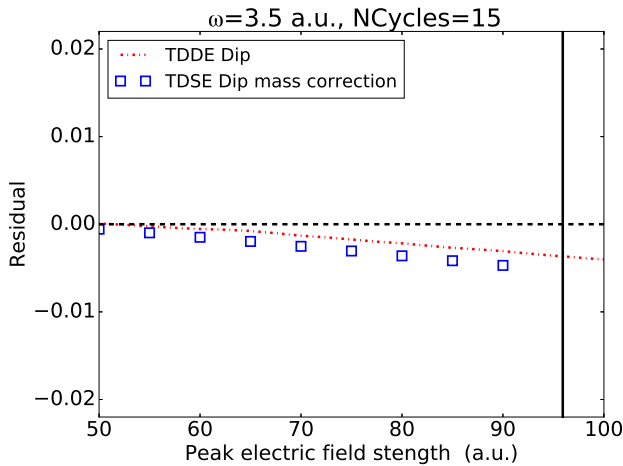}
\caption{(color online). Residual plots for the relativistic dipole calculation and the non-relativistic dipole calculation with the mass substituted as in Eq.~(\ref{RelMass}).}
\label{Residual_plot_masscorr}
\end{figure}

It is clear from Fig.~\ref{Residual_plot} that relativistic effects reduce the ionization probability. It would seem reasonable to assume that this reduction is related to the increased inertia of the electron. In order to test this assumption we have performed non-relativistic calculations in the dipole approximation in which the electron mass has been substituted with the relativistic mass of a classical free electron,

\begin{equation}
\label{RelMass}
m \rightarrow
\frac{m}{\sqrt{1-(v(t)/c)^2}} \quad \text{where} \quad
v(t) = \frac{e}{m} A(t) \quad .
\end{equation}
The difference between the corresponding ionization probability and the one obtained without mass shift is shown in Fig.~(\ref{Residual_plot_masscorr}) -- along with the TDDE results. Indeed we find that the ``relativistic substitution'', Eq.~(\ref{RelMass}), shifts the ionization probability downwards. Moreover, although our model overestimates the discrepancy somewhat,  the ionization probability obtained by the model behaves in a manner very similar to the truly relativistic calculations.

\begin{figure}[!h]
\includegraphics[width=0.50\textwidth]{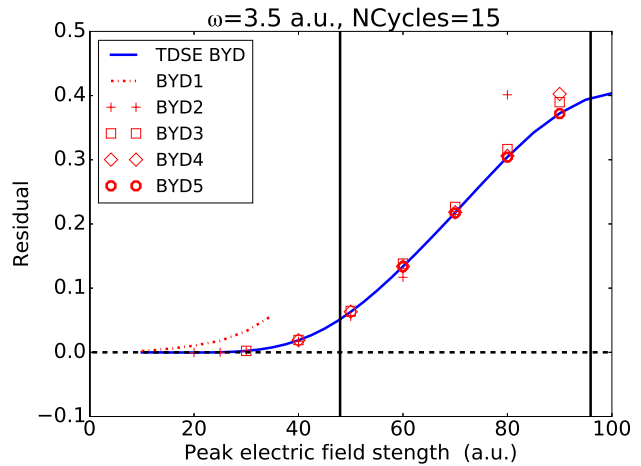}
\caption{(color online). Difference between $P_\mathrm{ion}$ computed within and beyond the dipole approximation for TDSE and TDDE in respective case.}
\label{BYD-dip_residual}
\end{figure}

This is a strong indication that the dominating relativistic effects for the fields considered here originate from dipole interaction. In support of this, Fig.~\ref{BYD-dip_residual} shows the difference between calculations performed within and beyond the dipole approximation -- both for the TDSE and the TDDE. It is seen that the corrections to the dipole approximation in the two frameworks do in fact coincide. Although it is hard to judge from Fig.~\ref{BYD3-5} alone if $n_\text{trunc} = 5$ is sufficient for $E_0 = 90$~a.u., the agreement in Fig.~\ref{BYD-dip_residual} provides strong
support thereof.

% % % % % % % % % % % % % % % % % % % % % % % % % % % % % % % % % % % %
% % % % % %    Results and discussion - subsection 3    % % % % % % % %
% % % % % % % % % % % % % % % % % % % % % % % % % % % % % % % % % % % %

\subsection{Dealing with the negative energy states}
\begin{figure}[!h]
\center
 \includegraphics[width=0.50\textwidth]{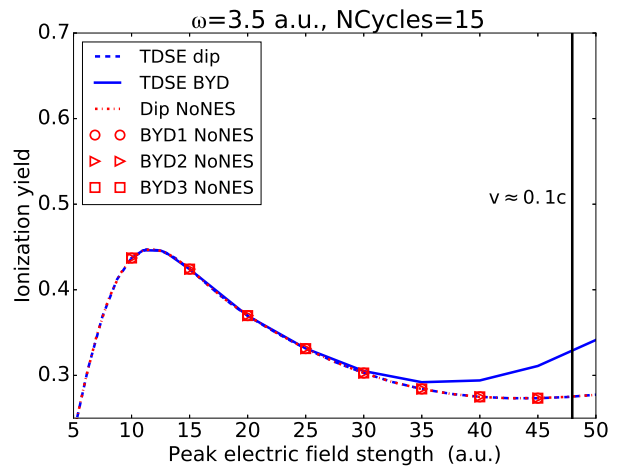}
\caption{(color online). A comparison of $P_\mathrm{ion}$ for the TDSE and the TDDE calculated with and without the time-independent negative-energy states in the relativistic basis. With this limitation, none of the simulations are able to provide any correction to the dipole approximation.}
\label{NoNES}
\end{figure}

In Ref.~\cite{Selsto2009} it was shown that exclusion of the time-{\em independent} negative-energy states from the propagation basis removed all effects beyond the dipole approximation. Here we show that this conclusion seems to hold regardless of the value of $n_\mathrm{trunc}$ in Eq.~(\ref{TaylorExpansion}). In Fig.~\ref{NoNES} we present  results from solving the TDDE without the time independent negative energy states, i.e., with those eigenstates of $H_0$ corresponding to negative eigenenergies excluded from the basis set, for $n_\mathrm{trunc} \leq 3$. It is seen that all these different Hamiltonians now predict the same $P_\mathrm{ion}$. 

In Sec.~\ref{PropAndProj} it was argued that the projection onto time-dependent (dynamic) positive energy states, i.e., eigenstates of the dynamical Hamiltonian $H(t)$, was adequate as long as the fields are not within the pair production regime.
This should however not be taken for granted if the propagator is evaluated within a Krylov subspace, cf. Eq.~(\ref{KrylovSpace}), since the obtained spectrum is not the same as the true time dependent spectrum of $H(t)$. 
To test this, two simulations for the TDDE BYD5, with $E_0 = 70$~a.u. and $80$~a.u., 
were performed both with and without the projection imposed within the Krylov subspace, cf. Eqs.~(\ref{PropagatorProj}, \ref{KrylovSpace}).
As it turns out, the only difference found between these calculations was slightly different convergence properties in the time step, which indicates that even in the approximate propagation method the arguments made in Sec.~\ref{PropAndProj} should hold.

% % % % % % % % % % % % % % % % % % % % % % %
% % % % % %    Conclusions    % % % % % % % %
% % % % % % % % % % % % % % % % % % % % % % %

\section{Conclusion}
\label{Conclusion}
We have solved the time-dependent Dirac equation for a hydrogen atom exposed to extreme laser pulses. Upon comparison with the non-relativistic counterpart, i.e., the Schr{\"o}dinger equation, it was found that effects beyond the dipole approximation are more complicated to incorporate correctly in the relativistic framework. Whereas first order space-dependent corrections to the vector field are sufficient for the TDSE, the TDDE demands an expansion to at least third order to even reproduce the non-relativistic results below the relativistic regime.
With increasing field strengths the demand for higher order corrections increases even further with the need of a fifth-order space-dependent correction when the quiver velocity is $v_\mathrm{quiv} \approx 0.19c$.

Emerging relativistic corrections are found in the ionization yield $P_{\text{ion}}$ for the test cases starting at $v_\mathrm{quiv} \approx 0.17c$. Both within and beyond the dipole approximation the relativistic effects give a lower $P_\mathrm{ion}$ suggesting increased stabilizing effect against ionization. It was demonstrated that this shift could be explained by the electron's increased relativistic inertia described already \textit{within} the dipole approximation.

To increase the field strength further and reveal stronger relativistic effects, possibly also relativistic corrections beyond the dipole approximation, the present form of the light-matter interaction in the Dirac equation is not suitable due to the demand of increasingly higher order corrections. A better approach is to look for a transformation of the Hamiltonian to a form where the consistent inclusion of higher order multipoles of the electromagnetic field is easier to achieve. Such a transformation  will be presented in a forthcoming paper.

% % % % % % % % % % % % % % % % % % % % % % % % % %
% % % % % %    Acknowledgements     % % % % % % % %
% % % % % % % % % % % % % % % % % % % % % % % % % %

\section*{Acknowledgments}
The authors would like to thank prof. M. F{\o}rre for fruitful discussions and for providing us with data for comparison. Significant development of the TDSE and TDDE codes were done at resources provided by the Swedish National Infrastructure for Computing (SNIC) HPC2N and NSC. We especially would like to thank Åke Sandgren at HPC2N for assistance in this development. The Norwegian
Metacenter for Computational Science (UNINETT Sigma2) is acknowledged for providing computational resources for carrying out the calculations (Account number NN9417K). Financial support by the Swedish Research Council (VR), Grant No. 2012-3668 is gratefully acknowledged. We also acknowledge support for our collaboration through the {\it Nordic Institute for Theoretical Physics} (Nordita) and from the research group {\it Mathematical Modelling} at Oslo and Akershus University College of Applied Sciences.

% % % % % % % % % % % % % % % % % % % % % % % % % %
% % % % % %       Bibliography      % % % % % % % %
% % % % % % % % % % % % % % % % % % % % % % % % % %

\end{document}